# Copyright Disclaimer

For academic and educational purposes only.

**To access the final version of this paper, please refer to:** 10.21314/JOIS.2023.003

**APA**
Ma, Y, et al. (2023). The realized local volatility surface. The journal of investment strategies, 12(1).

**Chicago Style**
Ma, Yuming, et al. "The Realized Local Volatility Surface." The Journal of Investment Strategies 12, no. 1 (2023).

**MLA**
Ma, Yuming, et al. "The Realized Local Volatility Surface." The Journal of Investment Strategies 12.1 (2023).

# Realized Local Volatility Surface

Yuming MA[1*], Shintaro SENGOKU[2], Kazuhide NAKATA[1]

June 28, 2023

[1]Department of Industrial Engineering and Economics, School of Engineering, Tokyo Institute of Technology, 2-12-1 Ookayama, Meguro-ku, Tokyo 152-8552 Japan E-mail: *ma.y.ae@m.titech.ac.jp, nakata.k.ac@m.titech.ac.jp

[2]Department of Innovation Science, School of Environment and Society, Tokyo Instiute of Technology, 2-12-1 Ookayama, Meguro-ku, Tokyo 152-8552 Japan E-mail: sengoku.s.aa@m.titech.ac.jp

**Abstract**

Estimation of the distribution of realized volatility is vital to options trading strategy formulation, as well as other portfolio risk management purposes. The main contribution of this article is to propose a Bayesian nonparametric estimation method to reconstruct a counterfactual generalized Wiener measure from historical price data. To reach this, a Stick-Breaking Gaussian Mixture Model (SB-GMM) has been applied to compensate for the return distribution over both time and price dimensions. For model paramters fitting, Hamilton monte-carlo (HMC) is applied to obtain SB-GMM paramters maximizing a posteriori probability (MAP method). Once a posteriori distribution is attained, we draw samples from it to compute the standard deviation and construct the realized local volatility surface at a 95% credibility intervals (CI). Thirdly, a numerical experiment result using ticker-level high-frequency data of Tesla, Inc. (Ticker: TSLA) to construct counterfactual realized local volatility surface is given. Finally, we addressed a possible usage that realized volatility surface compensates hedge-used implied volatility for managing risks of abrupt movement in the underlying price.

*Keywords*: Realized Volatility, Volatility Trading, Local Volatility, Counterfactuals Construction, Bayesian Nonparametric

## 1 Introduction

The objective of this paper is to construct an aligned comparison criterion over the implied volatility surface with respect to the real underlying price volatility process. The difference between implied volatility $\sigma_{imp}$ and realized volatility $\sigma_{real}$ is one of the essential drivers of profit and loss (PnL) generated from an options trading strategy(Neuberger 1994; Bakshi and Kapadia 2003; Sinclair 2013; Sepp 2010; Hull and Sinclair 2022). Here, implied volatility $\sigma_{imp}$ inversely computed from options market quoted price, and realized volatility $\sigma_{real}$ refers to the historical underlying asset price dynamics. The ofen-overlooked part of the comparison, however, is that $\sigma_{real}$ is a stochastic process while $\sigma_{imp}$ has a surface as the options differing. A discrepancy occurs when traders are evaluating their portfolio risks and conducting hedging activities. To compensate the discrepancy, model captures interactions between the underlying price process and its volatility process gaining popularity among practitioners. Besides models taking forms of stochastic differential equations (SDEs), e.g., Heston model(Heston 1993) or SABR model(Hagan et al. 2002) etc. Previous reseaerch(Heston and Nandi 1998) utilizes GARCH processes to describe estimations of stochastic underlying conditional variance processes and recorded an error-reduced result comparing with the result from the Black-Scholes model which takes historical implied volatility as inputs. However, as for a trading perspective, acknowledging an credible interval of the underlying price volatility over time and price dimensions is a more significant concern. Hence, we apply a Bayesian nonparametric fashioned model to build credible counterfactual intervals remedying the implications of would-probably-have-happened historical market data via imitating the contributions to the underlying log-return process from various market participants behaviors.

The rest of the article is organized as follows. Section 2 exhabits the relationship options trading PnL and implied and realized volatility under a generalized stochastic volatility setting and implied volatility surfaces. In section 3 we refer to Dupire's local volatility (Dupire 1994; Derman and Kani 1994) then we discuss the distinction with our Bayesian nonparametric estimation approch. Section 4 formulates our model in mathmatical details. In general, our approch is a probabilistic generative multi-dimensional mixing Gaussian process model. The mixing weights in our model is designed to be a probability measure in order to construct random probability measures (RPMs) to fit underlying market quotes dynamics. Section 5 presents the optimization method named Hamilton Monte-Carlo method to effectively generate a set of generative model parameters for the purpose of maximizing a posteriori probability (MAP). Section 6 illustrates a numerical experiment based on ticker-level high-frequency market data of Tesla, Inc. (Ticker: TSLA) to estimate a realized volatility surface. Section 7 is for discussions and conlusions.

## 2 Relationship between PnL and volatility

We firstly show a PnL decomposition for dynamic delta hedged option trading strategy as a way to clarify the discrepancy between ex-ante realized variance $\sigma_{real}^2$ and implied variance $\sigma_{imp}^2$ is the source of trading PnL. Assuming there are stochastic processes in the underlying price process pace$\{S\}$, the short-rate process space $\{r\}$ and the volatility process space $\{\sigma\}$ are Ito processes defined on a standard filtration equipped complete probability space $(\Omega, \mathcal{F}, \mathbb{P}; \{\mathcal{F}_t\}_{t\in[0,T]})$. These processes are depicted by a generalized system of SDEs:

$$\begin{cases} dS_t = r(t, S_t)dt + \sigma(t, S_t)dW_t^S \\ d\sigma_t^2 = \kappa(t, \sigma_t)dt + \xi(t, \sigma_t)dW_t^\sigma \\ dr_t = \theta(t, r_t)dt + \zeta(t, r_t)dW_t^r \end{cases} \qquad (2.1)$$

with a correlation structure

$$\begin{pmatrix} dW_t^S \\ dW_t^\sigma \\ dW_t^r \end{pmatrix} = \begin{pmatrix} 1 & 0 & 0 \\ \rho_{s\sigma_t} & \sqrt{1-\rho_{s\sigma_t}^2} & 0 \\ \rho_{sr_t} & \rho_{\sigma r_t}\sqrt{1-\rho_{sr_t}^2} & \sqrt{1-\rho_{sr_t}^2-\rho_{\sigma r_t}^2} \end{pmatrix} \begin{pmatrix} dZ_t^S \\ dZ_t^\sigma \\ dZ_t^r \end{pmatrix} \quad (2.2)$$

where $dZ_t^S, dZ_t^\sigma, dZ_t^r$ are respectively independent Wiener process incrementals and $\{\rho\}$ are instantaneous correlation coefficients.

Since an option of underlying asset, denoted as $V_t(\{S\}) \in C^2(\mathbb{R}^N)$, we apply Ito's lemma (Kusuoka 2020) to the market-to-market option value $V_{t_0+\tau}$ at $t_0+\tau$ w.r.t the stochastic processes solved from Eq. (2.1), we have options value process

$$V_{t_0+\tau} = V_{t_0} + \int_{t_0}^{t_0+\tau} \begin{pmatrix} \frac{\partial V_u}{\partial u} & \frac{\partial V_u}{\partial r_u} & \frac{\partial V_u}{\partial S_u} & \frac{\partial V_u}{\partial \sigma_u} \end{pmatrix} \begin{pmatrix} du \\ dr_u \\ dS_u \\ d\sigma_u \end{pmatrix}$$

$$+ \int_{t_0}^{t_0+\tau} \begin{pmatrix} du & dr_u & dS_u & d\sigma_u \end{pmatrix} \begin{pmatrix} \frac{\partial^2 V_u}{\partial u^2} & \frac{\partial^2 V_u}{\partial u \partial r_u} & \frac{\partial^2 V_u}{\partial u \partial S_u} & \frac{\partial^2 V_u}{\partial u \partial \sigma_u} \\ \frac{\partial^2 V_u}{\partial r_u \partial u} & \frac{\partial^2 V_u}{\partial r_u^2} & \frac{\partial^2 V_u}{\partial r_u \partial S_u} & \frac{\partial^2 V_u}{\partial r_u \partial \sigma_u} \\ \frac{\partial^2 V_u}{\partial S_u \partial u} & \frac{\partial^2 V_u}{\partial S_u \partial r_u} & \frac{\partial^2 V_u}{\partial S_u^2} & \frac{\partial^2 V_u}{\partial S_u \partial \sigma_u} \\ \frac{\partial^2 V_u}{\partial \sigma_u \partial u} & \frac{\partial^2 V_u}{\partial \sigma_u \partial r_u} & \frac{\partial^2 V_u}{\partial \sigma_u \partial S_u} & \frac{\partial^2 V_u}{\partial \sigma_u^2} \end{pmatrix} \begin{pmatrix} du \\ dr_u \\ dS_u \\ d\sigma_u \end{pmatrix}$$
(2.3)

From the property of quadratic variation, we deduce

$$\begin{pmatrix} (dt)^2 & dt\,dr_t & dt\,dS_t & dt\,d\sigma_t \\ dr_t\,dt & (dr_t)^2 & dr_t\,dS_t & dr_t\,d\sigma_t \\ dS_t\,dt & dS_t\,dr_t & (dS_t)^2 & dS_t\,d\sigma_t \\ d\sigma_t\,dt & d\sigma_t\,dr_t & d\sigma_t\,dS_t & (d\sigma_t)^2 \end{pmatrix}$$

$$= \begin{pmatrix} 0 & 0 & 0 & 0 \\ 0 & \zeta_t^2 dt & \sigma_t\zeta_t\alpha_t dt & \frac{1}{2}\zeta_t\sqrt{\langle \xi_t,\xi_t\rangle}\gamma_t dt \\ 0 & \sigma_t\zeta_t\alpha_t dt & \sigma_t^2 dt & \frac{1}{2}\sigma_t\sqrt{\langle \xi_t,\xi_t\rangle}\beta_t dt \\ 0 & \frac{1}{2}\zeta_t\sqrt{\langle \xi_t,\xi_t\rangle}\gamma_t dt & \frac{1}{2}\sigma_t\sqrt{\langle \xi_t,\xi_t\rangle}\beta_t dt & \frac{1}{4}\langle \xi_t,\xi_t\rangle dt \end{pmatrix} \quad (2.4)$$

$$\alpha_t = \rho_{sr_t}, \beta_t = \rho_{s\sigma_t}, \gamma_t = \rho_{sr_t}\rho_{s\sigma_t} + \rho_{\sigma r_t}\sqrt{(1-\rho_{sr_t}^2)(1-\rho_{s\sigma_t}^2)}$$

where $\langle,\rangle$ is the quadratic variation operator.

Substituting Eq. (2.4) into Eq. (2.3), we have

$$\triangle V_{t_0} = V_{t_0+\tau} - V_{t_0} \quad (2.5)$$

$$= \int_{t_0}^{t_0+\tau} \frac{\partial V_u}{\partial u} du$$

$$+ \int_{t_0}^{t_0+\tau} \frac{\partial V_u}{\partial S_u} dS_u + \frac{1}{2}\int_{t_0}^{t_0+\tau} \frac{\partial^2 V_u}{\partial S_u^2} \sigma_u^2 du$$

$$+ \int_{t_0}^{t_0+\tau} \frac{\partial V_u}{\partial \sigma_u} d\sigma_u + \frac{1}{8}\int_{t_0}^{t_0+\tau} \frac{\partial^2 V_u}{\partial \sigma_u^2}\langle \xi_u,\xi_u\rangle du + \frac{1}{2}\int_{t_0}^{t_0+\tau} \frac{\partial^2 V_u}{\partial S_u \partial \sigma_u}\sigma_u\sqrt{\langle \xi_u,\xi_u\rangle}\beta_u du$$

$$+ \int_{t_0}^{t_0+\tau} \frac{\partial V_u}{\partial r_u} dr_u + \frac{1}{2}\int_{t_0}^{t_0+\tau} \frac{\partial^2 V_u}{\partial r_u^2} \zeta_u^2 du$$

$$+ \int_{t_0}^{t_0+\tau} \frac{\partial^2 V_u}{\partial S_u \partial r_u}\sigma_u\zeta_u\alpha_u du + \frac{1}{2}\int_{t_0}^{t_0+\tau} \frac{\partial^2 V_u}{\partial r_u \partial \sigma_u}\zeta_u\sqrt{\langle \xi_u,\xi_u\rangle}\gamma_u du$$

where the first order greeks are:

$$Theta_t := \frac{\partial V_t}{\partial t} \qquad Rho_t := \frac{\partial V_t}{\partial r_t} \qquad Delta_t := \frac{\partial V_t}{\partial S_t} \qquad Vega_t := \frac{\partial V_t}{\partial \sigma_t}$$

and the commonly used second order greeks are:

$$Gamma_t := \frac{\partial^2 V_t}{\partial S_t^2} \qquad Vanna_t := \frac{\partial^2 V_t}{\partial S_t \partial \sigma_t} \qquad Volga_t := \frac{\partial^2 V_t}{\partial \sigma_t^2} \qquad Vera_t := \frac{\partial^2 V_t}{\partial r_t \partial \sigma_t}$$

Traders may use specific type of greeks generated from their own trading systems, however, generally the better PnL performance for a delta neutral strategy is closely related to the accurate estimate of future volatility (Bakshi and Kapadia 2003; Sinclair 2013; Bennett 2014).

Following an ordinary Black-Scholes-Merton model setting with convenience yield rate $q_t$, we have

$$(dS_t)^2 = \sigma_{real,t}^2 S_t^2 dt \qquad (2.6)$$

$$Theta_t = -\frac{1}{2} Gamma_t \sigma_{imp,t}^2 S_t^2 dt - \frac{r_t}{T-t} Rho_t - \frac{q_t}{T-t} \frac{\partial V_t}{\partial q_t} \qquad (2.7)$$

Taking Eq. (2.6) and Eq. (2.7) into Eq. (2.5) and meanwhile eliminating the delta term due to its neutrality under dynamically delta hedging strategy, we have monoperiod PnL:

$$\triangle V_{t_0} = \frac{1}{2} \int_{t_0}^{t_0+\tau} \frac{\partial^2 V_u}{\partial S_u^2}(\sigma_{real,u}^2 - \sigma_{imp,u}^2)S_u^2 du + \frac{1}{2} \int_{t_0}^{t_0+\tau} \frac{\partial^2 V_u}{\partial S_u \partial \sigma_u} \sigma_{real,u} \xi(u, \sigma_{imp,u}) S_u \beta_u du \qquad (2.8)$$

$$+ \int_{t_0}^{t_0+\tau} \frac{\partial V_u}{\partial \sigma_u} d\sigma_u + \frac{1}{8} \int_{t_0}^{t_0+\tau} \frac{\partial^2 V_u}{\partial \sigma_u^2} \xi^2(u, \sigma_{imp,u}) du$$

$$+ \int_{t_0}^{t_0+\tau} (1 - \frac{r_t}{T-u})\frac{\partial V_u}{\partial r_u} dr_u + \frac{1}{2} \int_{t_0}^{t_0+\tau} \frac{\partial^2 V_u}{\partial r_u^2} \zeta_u^2 du$$

$$+ \int_{t_0}^{t_0+\tau} \frac{\partial^2 V_u}{\partial S_u \partial r_u} \sigma_u \zeta_u \alpha_u du + \frac{1}{2} \int_{t_0}^{t_0+\tau} \frac{\partial^2 V_u}{\partial r_u \partial \sigma_u} \zeta_u \xi(u, \sigma_{imp,u}) \gamma_u du$$

Eq. 2.8 exhibits that PnL from a delta-neutral option trading strategy is sourced on the discrepency process of the realized-in-future variance process $\sigma_{real,t}^2$ and the implied variance process $\sigma_{imp,t}^2$. The former is the actual dynamics of the underlying after the strategy starts running, while the latter is implied from options market quotes and is the expected variance under an equivalent martingale measure(Ahmad and Wilmott 2005; Derman and Miller 2016).

Here we provide more in-depth intuition for delta-neutral option trading strategy. Eq. 2.8 shows the gamma exposure and the vanna exposure are corely related to the underlying price dynamics in future. Keeping aggregated portfolio delta being neutralized, trading PnLs are realized incrementally by increasing or liquidating gross positions multiple times in underlying assets. Gross position is simply the absolute value of signed position, e.g. short selling takes a negative sign. If a portfolio is with positive aggregated gamma, any upward movement in the underlying price contrubutes a positive delta increment to the portfolio, then delta-neutral trading strategy policy will require a negative delta increment, i.e. a selling on underlying asset inventories, at a certain stopping time for re-neutralization, *vice versa*. Note that there is a relationship, shown by Eq. 2.6, between realized variance $\sigma^2_{real,(t,t+\tau]}$ in future and return in underlying asset $R_{(t,t+\tau]} := \frac{S_{t+\tau}-S_t}{S_t} \approx \log \frac{S_{t+\tau}}{S_t}$ during a discretized delta re-neutralization time interval $(t, t+\tau]$.

$$\sigma^2_{real,(t,t+\tau]} = \frac{1}{\tau}\frac{(S_{t+\tau}-S_t)^2}{S_t^2} = \frac{1}{\tau}R^2_{(t,t+\tau]} \approx \frac{1}{\tau}\log^2\frac{S_{t+\tau}}{S_t} =: r_{(t,t+\tau]} \quad (2.9)$$

In order to generate profits on gamma while eliminating most of market directional risk, it is essential for a dynamic delta hedging strategy manager to keep aggregated gamma sharing identical sign with the discrepency of annualized Eq. 2.9 and implied variance of each option within the portfolio thoughout a sufficient number of delta re-neutralization time intervals, as it has been exhibited in Eq. 2.8. Nevertheless, market directional risk can remains explicitly or implicitly present in vanna, vega, and other greeks due to market interactions between volatility surface and the underlying price dynamics (Derman 1999; Daglish, Hull, and Suo 2007; Bergomi 2009).

Therefore, it would be versatile to estimate the expected underlying variance at an arbitrary given time and underlying asset price. The given time and underlying price are not necessarily recorded in the history, in fact what we are motivated to contribute in this paper is a Stick-Breaking Gaussian Mixture Model (SB-GMM) to construct a credible counterfactual local realized volatility surface. The local realized volatility surface contains conditional expectations of annualized volatility under the physical probability measure $\mathbb{P}$ in underlying price path space. To some extent, it is similar to Dupire's local volatility function. Before presenting our model, the next section describes the differences with Dupire's local volatility model.

## 3  Local Volatility and Path Space

Dupire's local volatility model (Dupire 1994; Derman and Kani 1994) is an alternavtive manner to be compatible with observed volatility smiles(skewnesses) contrasting with stochastic volatility models. Dupire, Derman and Kani developed the idea that a risk-neutral probability density can be derived from non-arbitrage market prices of all European options(Breeden and Litzenberger 1978). They introduced a determinestic instantaneous diffusion function drived from makret quoted options prices to extend the original Black-Scholes model.

$$dS_t = \mu S_t dt + \sigma^{LV}(T, K; S_t, t)S_t dW_t \quad (3.1)$$

where the local volatility function $\sigma^{LV}(T, K; S_t, t)$ is calimed as an expectation of possible instantaneous volatility at expiry $\sigma_T$ conditional on the underlying price at expiry $S_T$ being exactly equal to the strike price $K$. The local volatility function is specified under given parameters that are current underlying price $S_t$ and market prices of all European call options $C := C(K, T; S_t, t)$ at expiry dates $\{T\}$ and a strike prices $\{K\}$.

$$\sigma^{LV}(K, T; S_t, t) := \sqrt{\mathbf{E}^{\mathbb{Q}}[\sigma_T^2 | S_t, S_T = K]}$$
$$= \sqrt{\frac{\frac{\partial C}{\partial T} + [r_t(T) - d_t(T)][K \frac{\partial C}{\partial K} - C]}{\frac{1}{2} K^2 \frac{\partial^2 C}{\partial K^2}}} \qquad (3.2)$$

Note that Dupire's local volatility does not present underlying volatility process in the real world. Technically Dupire's local volatility is a risk-neutral conditional expectation of the instantaneous volatility forward price corresponded with the underlying asset.

On the contrary, our method aims to estimate an ex-post expected realized local volatility function $\sigma_{local}(s, t)$ given an arbitrary price $s$ and time $t$, which is

$$\sigma_{local}(s, t) := \sqrt{\mathbf{E}_t^{\mathbb{P}}[\langle S_t, S_t \rangle | t, S_t = s]} = \sqrt{\int_{\Omega_t} \sigma^2(S_t, t; \boldsymbol{\theta}_t) h_t(\boldsymbol{\theta}_t | S_t = s) d\boldsymbol{\theta}_t}$$
(3.3)

note that the probability measure $\mathbb{P}$ is conditional on $\{\mathcal{F}_t\}$, thus we take Eq. 3.3 as a proxy of the ex-ante realized-in-future volatility. We assume that a model to estimate realized local volatility $\sigma_{local}(s, t)$ is parameterized by $\boldsymbol{\theta}$. $\sigma^2(S_t, t; \boldsymbol{\theta}_t)$ is the instantaneous volatility of the underlying asset at time $t$ given a model paramter $\boldsymbol{\theta}_t$. $h_t(\boldsymbol{\theta}_t | S_t = s)$ is a posteriori probability density observing a particular fitted paramter $\boldsymbol{\theta}_t$ by given a condition that the underlying price $S_t = s$.

Our approach, SB-GMM, and its parameter calibration procedure, Hamilton Monte-Carlo, in the following sections are designed to find the most possible probability measure $\mathbb{P}$ among an abstract path space $(\Psi_T, \mathscr{B}(\Psi_T), \mathbb{P})$ on $[0, T]$. We consider the probability measure $\mathbb{P}$ in the form of the Wiener measure, but extend the Gaussian kernel into a mixed Gaussian kernel as to accommodate leptokurtic and negatively skewed market return distributions. $\Psi_T$ is the set of $\mathbb{R}^d$-valued continuous functions $R$ on time on $[0, T]$ with $R(0) = 0$. To be intuitive, $\Psi_T$ contains all possible returns measured during any arbitrary time interval. Specifically, taking the underlying return process as $R_{(t,t+\tau]}$, we define such a probability measure $\mathbb{P}$ for $0 = t_0 < t_1 < t_2 < \cdots < t_n$ and $\Lambda_1, \Lambda_2, \cdots, \Lambda_n \in \mathscr{B}(\Psi_T)$:

$$\mathbb{P}[\{R \in \Psi_T; R_{(t_0, t_1]} \in \Lambda_1, R_{(t_1, t_2]} \in \Lambda_2, \cdots, R_{(t_{n-1}, t_n]} \in \Lambda_n\}] \qquad (3.4)$$
$$= \int_{\Lambda_1} \int_{\Lambda_2} \cdots \int_{\Lambda_n} \prod_{i=1}^{n} \varphi(t_i - t_{i-1}, x_i, x_{i-1}) dx_1 dx_2 \cdots dx_n$$

where $\varphi(t_i - t_{i-1}, R_i, R_{i-1})$ is the probability density function for a certain value of the underlying return $R_i$ occurring by given a privious value $R_{i-1}$ over a given time interval $(t_i, t_{i-1}]$. With historical data, one can find a corresponded ex-post realized volatility process in addition to a return process. It is the realization from the probability measure $\mathbb{P}$, but there still remains other possible return processes that have not been realized in the past. To the purpose of compensating the other possible conterfactual paths, SB-GMM depicts $\varphi$ as a mixed

Gaussian measure of the underlying log-return process. In the subsequent article we use a linear model to represent the expectation values of those component measures that composed the mixed Gaussian measure in relation to time and price. Hamilton Monte-Carlo (HMC) method is applied to find a model parameter set $\theta$ to maximize a posteriori probability given observed historical return process. After we have a maxmized posteriori probability, we then draw log-return samples and compute their standard deviation and quantiles to estimate $\sigma_{local}(s,t)$ and its credible interval.

# 4 Stick-Breaking Gaussian Mixture Model (SB-GMM)

SB-GMM is a stick-breaking weigthed Gaussian mixture model. Its weights are drawn from the stick-breaking process(Griffin and Steel 2011), which ensures any SB-GMM is a random probability measure(RPM). We use it as a Bayesian likelihood measure for depicting the probability that log-returns occured within a certain time-price grid. A time-price grid partitions a joint region of time and price as it is shown in Fig. 4.1. We index time-price grids with $i \in 1, 2, \ldots, I, j \in 1, 2, \ldots, J$ respectively.

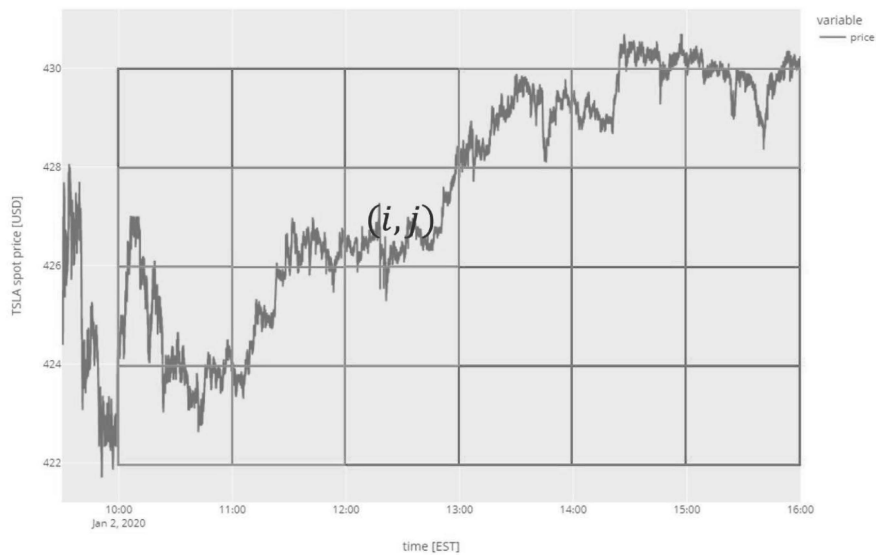

Figure 4.1: An instance of time-price grids draw from historical price data of TSLA. Each grid is indexed by $i, j$. The green colored grids represent areas where historical underlying price process passed through, while the red colored grids represent where was not. Then a Boolean mask is built to represent the path where red grids are 0 while green grids are 1 for further model fitting computation.

SB-GMM differs from normal Wiener process settings. For a Wiener process, its increments are generated from a Gaussian distribution with expectation 0 and variance $dt$. On the other hand, SB-GMM treats stochastic increments as are generated from multiple Gaussian distribution with various expectations $\mu_{i,j,k}$ and variance 1 as it is explained thereafter. With the stick-breaking wights, SB-GMMs combines multiple unit variance Gaussian distributions to a distribution which displays flexibility to higher-order moments. Due to Within any a time-price grid indexed by $i,j$, there is a set of records if historical underlying price data passed that grid (e.g. green grids in Fig. 4.1). If there was no historical underlying price data being recorded, the set is a null set (e.g. red grids in Fig. 4.1).

For details of model settings, we have a random variable $r_{i,j}$ standing for high-frequency log-return within a time-price grid indexed by $i,j$:

$$r_{i,j} := \sum_{n=2}^{N_{i,j}} log(\frac{S_n}{S_{n-1}}) = \sum_{n=2}^{N_{i,j}} log(S_n) - log(S_{n-1}) \quad (4.1)$$

and we assume that $r_{i,j}$ follows a probability measure SBGMM$_{i,j}$ realized from SB-GMM within the $i,j$ time-price grid.

$$r_{i,j} \sim \text{SBGMM}_{i,j}$$

$$\text{SBGMM}_{i,j} := \sum_{k=1}^{K} w_{i,j,k} \text{Normal}(\mu_{i,j,k}, 1) \quad (4.2)$$

$$w_{i,j,k} := \gamma_{i,j,k} \prod_{l=1}^{k-1} (1 - \gamma_{i,j,l})$$

$$\gamma_{i,j,k} \sim \text{Beta}(1, \alpha_{i,j}) \quad iid \quad (4.3)$$

There is a necessity for SBGMM$_{i,j}$ being a probability measure, hence the weights should be surely sum up to 1. Stick-breaking process, Eq. 4.3, is such a process to generate a series of a series of random variables summing to 1. It is a realization form of a stochastic process family named Dirichlet process, and random variables in a stick-breaking process path shows power law (Ferguson 1973; Teh 2010; Griffin and Steel 2011). There is an advantage of adapting a power law weighted mixed Gaussian distribution that stick-breaking weights make the final variance largely influenced by the squared expectation of the most weighted component Gaussian distribution by our SB-GMM setting, Eq. 4.4. To show this mathmatically, we provide expresions for the final expectation and variance of any a mixed Gaussian distribution as below, whilst $\sigma_i = 1$ in SB-GMM,

$$E[X] = \sum_{i=1}^{n} w_i \mu_i$$

$$E[(X - E[X])^2] = \sum_{i=1}^{n} w_i(\sigma_i^2 + \mu_i^2) - \sum_{i=1}^{n} w_i \mu_i \quad (4.4)$$

we are capable to leverage SBGMM to capture the most powerful stochastic flow keeping other weaker ones without losing generality. Further, we provide an intuitive explaination under financial market context. SBGMM describes multiple stochastic flows generated by multiple trading venues, some with higher explanatory power for underlying asset volatility, and their influence is stronger, while the rest are weaker and do not "drive the market".

As it is shown below, we next add the dependence on price to component measures of SBGMM. We can indicate that different trading venues will react differently when the price moves to a certain level. Some will reject the lower price because it will leave them inadequate in capital. Some will reject higher prices because it will leave them with too much new inventory to dispose of in a risk-managed manner. With SBGMM, we have more flexibility in expressing the market returns distribution. Expectations of component measures $\mu_{i,j,k}$ are estimated via a series of multi-factor linear regressions. We initialize Eq. (4.5) with independently drawn samples from standard normal distributions for coefficients.

$$\mu_{i,j,k} := \text{time\_effect}_{i,k} \times t_i + \text{price\_effect}_{j,k} \times log(S_i) + \text{alpha}_k + \epsilon_{i,j,k} \quad (4.5)$$

$$\begin{pmatrix} \text{time\_effect}^{(0)}_{i,k} \\ \text{price\_effect}^{(0)}_{j,k} \\ \text{alpha}^{(0)}_k \\ \epsilon_{i,j,k} \end{pmatrix} \sim \begin{pmatrix} \text{Normal}(0,1) \\ \text{Normal}(0,1) \\ \text{Normal}(0,1) \\ \text{Normal}(0,1) \end{pmatrix} \quad (4.6)$$

The optimization method to maximize a posteriori (MAP) is Hamiltonian Monte-Carlo method, which will be breifly introduced in section 5.

Fig. 4.1 shows that historical underlying price data solely occupy a part of the entire $(S_t, t)$ space due to the realization of the law of one price. Nevertheless, a local volatility function should be given across the whole time-price space. Wherefore a challenge on model parameters fitting rises that our loss function is merely meaningful w.r.t the time-price grids containing with historical data. We suggest filtering historical prices with a boolean mask to calibrate model parameters strictly for grids with historical prices.

Bayesian nonparametric estimation using MAP aims to find a model parameter set $\boldsymbol{\theta}$ as an possible solution of the optimization problem Eq. 4.7. The solution is found via minizing the inverse number of log poseriori $\mathbb{P}(\boldsymbol{\theta}|X)$, which equals to the right hand, the inverse number of the sum of likelihood SBGMM(X|$\boldsymbol{\theta}$) and generation priori $\Pi(\theta)$ of model parameters, because of Bayesian formula. Given an observation dataset $X$, we have

$$\min_{\boldsymbol{\theta}} - \log\left(\mathbb{P}(\boldsymbol{\theta}|X)\right) \propto \min_{\boldsymbol{\theta}} - \log\left(\text{SBGMM}(X|\boldsymbol{\theta})\right) - \log\left(\Pi(\theta)\right) \quad (4.7)$$

where the model paramter $\boldsymbol{\theta}$ contains all parameters $\boldsymbol{\theta}_{i,j,k}$ that describe a sepecific GMM component

$$\boldsymbol{\theta}_{i,j,k} := \begin{pmatrix} \text{time\_effect}_{i,k} \\ \text{price\_effect}_{j,k} \\ \text{alpha}_k \\ \alpha_{i,j} \end{pmatrix} \sim \begin{pmatrix} \text{Normal}(0,1) \\ \text{Normal}(0,1) \\ \text{Normal}(0,1) \\ \text{Beta}(1,1) \end{pmatrix} \quad iid \quad (4.8)$$

For any a GMM indexed by $i, j$, we have an observation log-return dataset $\boldsymbol{r}_{i,j}$. We take every data in $\boldsymbol{r}_{i,j}$ inputed into SBGMM$_{i,j}$ to obtain likelihood probability SBGMM($X|\boldsymbol{\theta}$) by a set of parameter $\boldsymbol{\theta}_{i,j}$ given.

$$\text{SBGMM} = \begin{pmatrix} \text{SBGMM}_{1,1} & \dots & \text{SBGMM}_{i,1} & \dots & \text{SBGMM}_{I,1} \\ \vdots & \ddots & \vdots & \ddots & \vdots \\ \text{SBGMM}_{1,j} & \dots & \text{SBGMM}_{i,j} & \dots & \text{SBGMM}_{I,j} \\ \vdots & \ddots & \vdots & \ddots & \vdots \\ \text{SBGMM}_{1,J} & \dots & \text{SBGMM}_{i,J} & \dots & \text{SBGMM}_{I,J} \end{pmatrix} \quad (4.9)$$

Eq. 4.9 shows a matrix form of our model across time-price grids. To optimize Eq. 4.7 we use Hadamard product of Eq. 4.9 and the Boolean mask.

$$\text{SBGMM}(X|\boldsymbol{\theta}) := \text{SBGMM} \circ \text{BooleanMask} \quad (4.10)$$

Eq. 4.10 is then maximized via HMC method to propose a set of possible model parameters.

## 5 Hamiltonian Monte-Carlo (HMC) Optimization

HMC is a Markov chain monte-carlo method combining Metropolis-Hasting algorithm and a Hamiltonian dynamics to accelerate convergence of random walks over possible parameter space(Brooks et al. 2011; Betancourt 2018). Given an unknown probability measure $\mathbb{P}$, e.g. the posteriori probability measure $\mathbb{P}(\boldsymbol{\theta}|X)$ in Bayesian statistics, our major interest is to find a parameter set $\boldsymbol{\theta} \in A$ where $A$ is an unkown solution set where every $\boldsymbol{\theta} \in \mathbb{R}^d$ inside is expected to maximize $\mathbb{P}(\boldsymbol{\theta}|X)$ as shown by Eq. 4.7 and Eq. 5.1.

$$\mathbb{P}(\boldsymbol{\theta}|X) \propto \mathbb{L}(X|\boldsymbol{\theta})\Pi(\boldsymbol{\theta}) \quad (5.1)$$

Maximizing the value of a probability function equals to minimize its inverse logarithm. Given an observation data $X$, we regard model parameter set $\boldsymbol{\theta}$ as the random variable. HMC, however, introduces Hamiltonian which is consistent with a set of mutually indenpendent functions, the potential energy $U(q)$ and the kinetic energy $K(p)$,

$$H(q, p) = K(p) + U(q) \quad (5.2)$$

and changes the unknown probability into a canonical ensemble probability measure via Radon–Nikodym theorem.

$$\mathbb{P}(\boldsymbol{\theta}|X) = \frac{1}{Z} e^{-\frac{H(q,p)}{T}} \quad (5.3)$$

where $q \in \mathbb{R}^d$ represents position and $p \in \mathbb{R}^d$ represents momentum. Variable pair $(q, p)$ represents a state of the system. System refers to any an iteration process over the parameter space to orient proposed paramter $\boldsymbol{\theta}^*$ to the solution set $A$. Hamiltonian measures system energy given a certain $(q, p)$. The canonical ensemble measures the probability for the system to be at the state $(q, p)$.

Instead of sampling from the unknown probability measure, HMC samples from the canonical ensemble probability measure. Iteration processes in HMC sampling comply with Hamiltonian dynamics and the law of conservation of energy.

$$\begin{cases} \frac{dq}{dt} = +\frac{\partial H(q,p)}{\partial p} = \frac{\partial K(p)}{\partial p} \\ \frac{dp}{dt} = -\frac{\partial H(q,p)}{\partial q} = -\frac{\partial U(q)}{\partial q} \end{cases} \quad (5.4)$$

where kinetic energy and potential energy have form of

$$K(p) := \frac{1}{2} p^T M^{-1} p \quad (5.5)$$
$$= -\log\left(\frac{1}{\sqrt{(2\pi)^d \det(M)}} e^{-\frac{1}{2} p^T M^{-1} p}\right) + \log\left(\frac{1}{\sqrt{(2\pi)^d \det(M)}}\right)$$

$$U(q) := -\log\left(\mathbb{L}(X|q)\right) - \log\left(\Pi(q)\right) \quad (5.6)$$

where $M$ is a positive definite symmetric matrix representing mass. The from of $K(p)$ here could be represented as an inverse number of logarithmic probability density of a $d$-dimensional multivariate normal distribution with $\mathbf{0}$ as its mean parameter and mass matrix $M$ as its covariance matrix and plus a positive constant number. Eq. 5.5 enables us to draw momentum $p$ from a standard multivariate normal distribution in HMC for optimization updates. On the other hand $U(q)$ is the inverse number of the right hand part of Eq. 5.1. $q$ is identical to $\boldsymbol{\theta}^*$ but is constrained to be within the HMC sampling iteration process.

Consequently Eq. 5.4 is a SDE according to the finding of Eq. 5.5 that momentum $p$ is independently and identically drawn from a standard multivariate normal distribution. The solution of Eq. 5.4 could be naïvely drived via the Euler-Maruyama method. Nevertheless, becuase discretization error largely impacts volume preservation property of Hamiltonian dynamics which leads to distortion in MAP estimation, second order discretization approximation called the leapfrog method is practically used to drive the solution process of Eq. 5.4. The leapfrog method is simillar to the 2nd-order Runge Kutta method, and it applies a half step update to ensure volume preservation property.

After the leapfrog method numerically solves the SDE Eq. 5.4 to raise a proposed update $(q^*, p^*)$, a algorithm will contingently decide if to accept the proposed update or remain in the present state $(q, p)$. The acceptance probability is

$$\min(1, e^{-H(q^*,p^*)+H(q,p)}) = \min(1, e^{-[U(q^*)-U(q)]-[K(p^*)-K(p)]}) \quad (5.7)$$

note that our goal is to find a $\boldsymbol{\theta}^*$ to maximize Eq. 5.3. The coloser the HMC sampling process position $q$ is to the unknown solution set $A$, the lower the acceptance probability comes out of Eq. 5. Thereupon HMC draws samples from $A$ which are expected to maximize a posteriori probability.

We alter Eq. 5.3 into below

$$H(q, p) = -T \log\left(Z\mathbb{P}(\boldsymbol{\theta}|X)\right) \quad (5.8)$$
$$\propto -T \log\left(\mathbb{L}(X|\boldsymbol{\theta})\right) - T \log\left(\Pi(\boldsymbol{\theta})\right) - T \log(Z) \quad (5.9)$$

Temperature $T$ and normalization denominator $Z$ are strictly positive. Therefore minimizing Eq. 5.8 can be approximately simplified to minimize

$$H(q,p) \propto -\log\left(\mathbb{L}(X|\boldsymbol{\theta})\right) - \log\left(\Pi(\boldsymbol{\theta})\right) \qquad (5.10)$$

note that it is what we expect in a MAP estimation as Eq. 4.7 shows.

$$\min_{\boldsymbol{\theta}} -\log\left(\mathrm{SBGMM}(X|\boldsymbol{\theta})\right) - \log\left(\Pi(\boldsymbol{\theta})\right) \Leftrightarrow \min_{(q,p)} H(q,p) \qquad (5.11)$$

To summarize, HMC develops a momentum-accelerated stochastic optimization method via drawing sampling processes as solutions of Eq. 5.4. Combine Eq. 5.2 with Eq. 5.5, Eq. 5.6, we take the expected value on both sides and have

$$\mathbf{E}[H(q,p)] = \mathbf{E}[-\log\left(\mathbb{L}(X|q)\right) - \log\left(\Pi(q)\right) + \frac{1}{2}p^T M^{-1} p] \qquad (5.12)$$

$$= \mathbf{E}[-\log\left(\mathbb{L}(X|q)\right) - \log\left(\Pi(q)\right)] \qquad (5.13)$$

$$\propto -\log\left(\mathrm{SBGMM}(X|\boldsymbol{\theta})\right) - \log\left(\Pi(\boldsymbol{\theta})\right) \qquad (5.14)$$

$$p \sim \mathrm{MultiNormal}(0, M) \quad iid$$

to minimize in order to have MAP fitted $\boldsymbol{\theta}^*$ of the solution set $A$. Fitted $\boldsymbol{\theta}^*$ is then given to the SB-GMM as its parameter to generate log-return samples to compute $\sigma_{local}(s,t)$ as shown in Eq. 3.3.

# 6 Numerical Experiment

As the samling density of historical underlying prices gains finer, we expect the sample mean and variance to be unbiased estimators w.r.t the population. Therefore, we leverage ticker-level high-frequency market transaction data. We used tensorflow for our implementation because it enables us to run the Bayesian nonparametric estimation on GPU, Nvidia RTX3090Ti, which is rather computational efficient. Time for the whole estimation and realized local volatility surface generation took indicatively 2 minutes over the entire intraday high-frequency data.

Fig. 6.1 shows the realized local volatility $\sigma_{local}(s,t)$ of TSLA on the trading day 2020-01-02 estimated via our SB-GMM Bayesian nonparameteric method. The time-price grids are constructed as following. For time dimension, we selected high-frequency trading data from 09:30 EST to 16:00 EST and resampled it by 5-minute frequency. On the price dimesion, we took the price space $[400, 450]$ and divide it into 10 partitions. We normalized data according to the price dimension and the time dimension. For normalization over time dimension, we took 09:30 as 0 and 16:00 as 1. To do so we divided the accumulated sum of differences of time between two next trade timestamps by the total 23,400 seconds within a trading day. For price, we changed the spot price $S_t$ of TSLA into a percentage number $s_t$.

$$s_t = \frac{S_t - MIN}{MAX - MIN} \qquad (6.1)$$

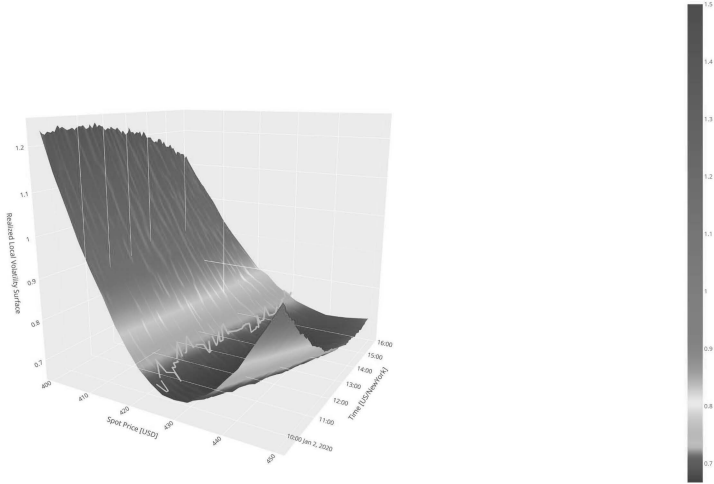

Figure 6.1: the expectation of the local volatility $\sigma_{local}(s,t)$ in decimal representation after 10,000 times simulations from the MAP estimated SB-GMM constructed by 5 componental Gaussian distributions. The highlight indicates the price of the underlying sampled at 5-minute intervals.

where $MIN$ is 400 and $MAX$ is 450.

We applied HMC method to update $\theta^*$ for 5000 times after a 1000-times updates as its burning length. In MCMC method, samples within the burning length are abandoned because the Markov chain may not converge to the stable stage and goal distribution within the length. For the appropriateness of the estimated model parameters, we find the acceptance rate of HMC was 75.26%. It means at 75.26% probability the model paramters has been successfully updated to be a better fitting model.

Realized local volatility estimation followed HMC application. We used expectation value of the lastest 100 SB-GMM parameter sets sampled via HMC method as the parameter of log-return generation model. We generated 100 times of 100 log-return samples from our fitted SB-GMM and computed their standard deviation within every time-price grid, which means 10,000 samples in total but with 100 different random seeds. Note we also adjust a 5-minute volatility to an annulized volatility. Fig. 6.1 shows the arithmetic mean of the standard deviation from each time of sampling. We can find that the realized local volatility function of TSLA on 2020/01/02 by SB-GMM shares the property of sticky-delta. Volatility in the areas that are near the underlying price varies little over time. Nevertheless, if one moves to the region where is far from the underlying price, the SB-GMM estimated realized local volatility shows higher results. We interpret this result that because of definition of realized local volatility, Eq. 3.3, the values on the surface are showing the expected volatility given a certain time and price. If the underlying price were 400 USD rather than 424 USD at 11:00 New York Time given the open price unchanged, a sharp 6% drop within 2 hours right after the market open would be recorded. In that counterfactual situation, the market expecation towards TSLA's volatilty should be largely elevated.

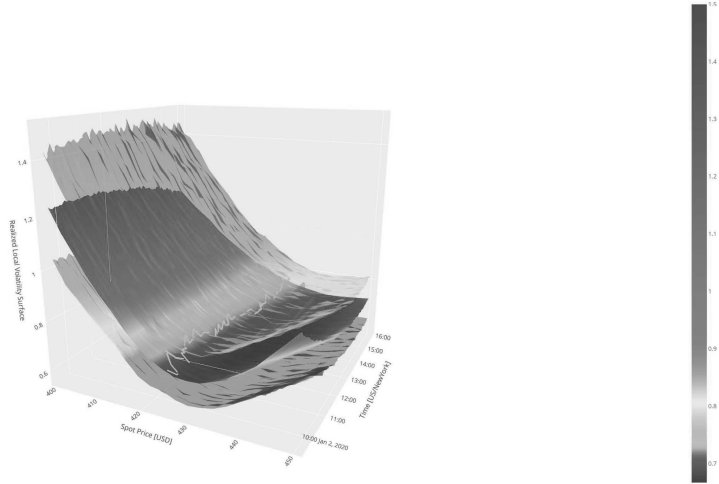

Figure 6.2: 95% credible interval of realized local volatility $\sigma_{local}(s,t)$ from fitted SB-GMM model. The space between the upper bound and the lower bound contains 95% of possible values of $\sigma_{local}(s,t)$. The highlight indicates the price of the underlying sampled at 5-minute intervals.

Fig. 6.2 shows the major advantage of our approach. We are able to draw estimation with expected error interval which surely contains possible results at the particular probability leavel. It is much intuitive and enables practioner to perform stress testing and scneario analysis and find solutions to deal with the worst possible result to their portfolio. In Fig. 6.2, we show the 95% credible interval (CI) of the SB-GMM estimated realized local volatility and the mean of it. CI is given by the 97.5% quantile and 2.5% quantile of the standard deviation.

Fig. 6.3 shows how SB-GMM estimated realized local volatility surfaces are different from implied volatility dynamics. We computed the implied volatility curves of OTM 1-day-to-expiry(1DTE) options at 5mins intervals. We choose 1DTE options for 2 reasons. Firstly, because near-expiry options are highly liquid in the US market, their market quotes can be considered as originating merely from the expectation of the volatility of the underlying price under the risk-neutral measure. Secondly, near-expiry options carry relatively large gamma around ATM strikes, and their prices are influenced by the underlying price dynamics at the most, therefore any dynamic delta hedging strategy manager would be prudent with them. SB-GMM estimated realized local volatility surfaces show about 20% higehr volatility than the 1DTE implied cruves exhibit near the underlying price. In addition to their range of values, the realized local volatility surfaces estimated by SB-GMM are with a very large curvature in the price direction. Recalling that our definition of the realized local volatility, Eq. 3.3, is the expecation of the quadratic variation given a certain price and time under the most possible probability measure $\mathbb{P}$, Eq. 3.4,among an abstract path space $(\Psi_T, \mathscr{B}(\Psi_T), \mathbb{P})$ on $[0,T]$, it is not a surprise due to the sticky-strike effect on the implied volatility surface. If a rapid movement happens, the implied volatility level would be squeezed up faster than the previous expectation.

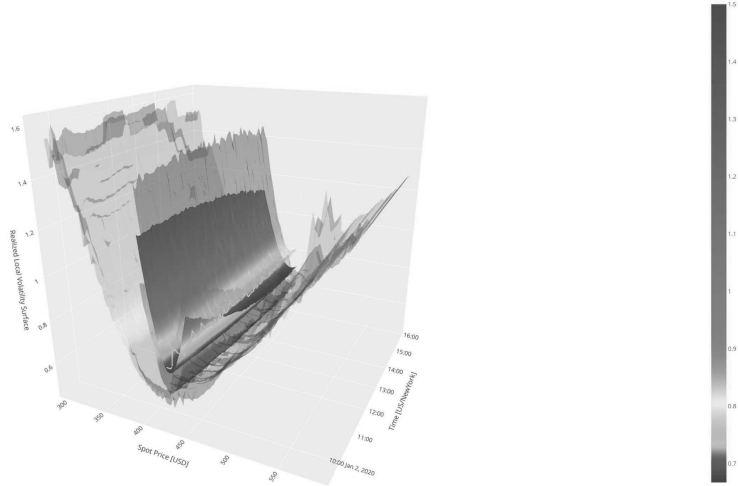

Figure 6.3: The SB-GMM-estimated realized local volatility surfaces (the three in the middle) are compared with the dynamics of the 1DTE implied volatility curve (the wider surface on the outside) for the entire trading day. Implied volatility curves are computed from TSLA OTM options mid-price by Newton-Raphson Method. Paramter settings: (1)Annual Trading Days: 252 Days, (2)26W T-Bill YTM: 1.53%, (3)Options Expiry: 2020-01-03 16:00 EST.

# 7 Discussion and Conclusion

In this section, we dicuss several insights could earn from the difference between SB-GMM estimated realized local volatility and the implied volatiliy curve dynamics. There are three major advantages of our approach: 1) it reconstructs a counterfactual generalized Wiener measure on the underlying price path space, 2) it gives an intuitive result that retains both the mean result and the estimation error interval, 3) it enables practitioners to perform stress testing and scenario analysis on the realized local volatility surface to better their risk managmenet procedure.

SB-GMM reconstructs a generalized Wiener measure from histroical market price data. We use Eq. 3.3 define our realized local volatility. It is under a probability measure $\mathbb{P}$ as Eq. 3.4. Probability measure $\mathbb{P}$ is one of the most possible generalized Wiener measure that is found by SB-GMM and HMC approach. Consequently the relized local volatility is a conditional expecation of the underlying price volatility, at a given time $t$ and a given underlying price $s$ leaving all other condition the same. For example, one can address a sudden market crush scenario, and find the counterfactual volatility on the estimated realized local volatility surface via our SB-GMM approach. Therefore, the numerically higher results in realized local volatility than the implied volatility on the both sides of the ATM anticipate the impact that would have been in implied volatility surface if a rapid movement or a large jump were being recorded.

To be more intuitive, imaging that TSLA spot price was 400USD instead of 424USD in history at 2020-01-02 11:00 EST would have happened. The estimated realized local volatility in Fig. 6.2 gives a CI of $[100\%, 147\%]$ and a mean of $123\%$ as its conditional expected annulized instantaneous volatility. This scenario means a rapid drop in TSLA price that a 6% movement occured within in the 2 hours after market open. Our SB-GMM approach enables practioners to take the result and test their portfolio if the volality level comes to $123\%$ at average and appears in the range of $[100\%, 147\%]$ at 95% probability.

For the model-wise contribution, we introduced stick-breaking process to constrain our GMM to generate random probability measures of log-return drived from the underlying price process without many detailed assumption in paramters. To efficiently have our SB-GMM fitted according to a high-frequency market data set given, we implemented Hamiltonian monte-carlo method and the whole Bayesian nonparamteric estimation workflow on GPU using tensorflow. We reached an indictivative estimation time within 2 minutes for the whole intraday data set. The numerical results from the same high-frequency data generated similar results, which shows our SB-GMM method have a certain degree of numerical consistency.

For the trading and risk management purpose, the counterfactual generalized Wiener measure constructed via our approach enriched practitioners toolbox for stress testing and scenario analysis. In detail, we suggested the realized local volatility surface is a conditional expectation set that represents instantaneous volatility of the underlying price process would be if a price and time set is given over the filtraion upto the given time. The large uncertainty in the conterfactual part of the estimated realized local volatility surface could be considered as the volatility that can be used in options pricer to evaluate the value and risk factors of a portfolio constituted by options under rapid movements. Specially for dynamic delta hedging strategy managers, in order to hedge their market directional risk under such a rapid moving secnario, the value on realized local volatility surface can be inputed to their greeks calculator as for ex-ante adjustments.

To summarize all, in this article we attempted to propose a new Bayesian nonparametric aiming at estimating realized local volatility function by using high-frequency financial market data. The originality of this article is that we proposed a Gaussian mixture model to construct a counterfactual generalized Wiener measure on the physical underlying price path space. Most of literatures or practices uses models from GARCH family to catch the statistical features merely along the time dimension(Hansen, Huang, and Shek 2010; Sharma and Vipul 2016). The SB-GMM and HMC as our estimation method gives practitioners two new flexibilities. The first is that they can obtain intuitively consistent expectation values and estimation errors from the estimated realized local volatility surface. The second is that they can perform stress tests and script analysis on the realized local volatility surface to determine the worst-case scenario to adjust each exposure in advance.

# DECLARATION OF INTEREST

The authors report no conflicts of interest. The authors alone are responsible for the content and writing of the paper.

# ACKNOWLEDGEMENTS

We thank the anonymous reviewers for their prudent and insightful remarks that improved the paper.